**Article type**: Research article

**Title**: Estimating the sample mean and standard deviation from commonly reported quantiles in meta-analysis

**Authors**:

Sean McGrath[1]

XiaoFei Zhao[1]

Russell Steele[2]

Brett D. Thombs[3-9]

Andrea Benedetti[1,5,6]

and the DEPRESsion Screening Data (DEPRESSD) Collaboration[10]

[1]Respiratory Epidemiology and Clinical Research Unit (RECRU), McGill University Health Centre, Montreal, Quebec, Canada

[2]Department of Mathematics and Statistics, McGill University, Montreal, Quebec, Canada

[3]Lady Davis Institute for Medical Research, Jewish General Hospital, Montreal, Quebec, Canada

[4]Department of Psychiatry, McGill University, Montreal, Quebec, Canada

[5]Department of Epidemiology, Biostatistics, and Occupational Health, McGill University, Montreal, Quebec, Canada

[6]Department of Medicine, McGill University, Montreal, Quebec, Canada

[7]Department of Psychology, McGill University, Montreal, Quebec, Canada

[8]Department of Educational and Counselling Psychology, McGill University, Montreal, Quebec, Canada

[9]Biomedical Ethics Unit, McGill University, Montreal, Quebec, Canada

[10]DEPRESSD Collaboration: Brooke Levis, McGill University, Montréal, Québec, Canada; Kira E. Riehm, Lady Davis Institute for Medical Research, Montréal, Québec, Canada; Nazanin Saadat, Lady Davis Institute for Medical Research, Montréal, Québec, Canada; Alexander W. Levis, McGill University, Montréal, Québec, Canada; Marleine Azar, McGill University, Montréal, Québec, Canada; Danielle B. Rice, McGill University, Montréal, Québec, Canada; Ying Sun, Lady Davis Institute for Medical Research, Montréal, Québec, Canada; Ankur Krishnan, Lady Davis Institute for Medical Research, Montréal, Québec, Canada; Chen He, McGill University, Montréal, Québec, Canada; Yin Wu, McGill University, Montréal, Québec, Canada; Parash Mani Bhandari, McGill University, Montréal, Québec, Canada; Dipika Neupane, McGill University, Montréal, Québec, Canada; Mahrukh Imran, Lady Davis Institute for Medical Research, Montréal, Québec, Canada; Jill Boruff, McGill University, Montréal, Québec, Canada; Pim Cuijpers, Vrije Universiteit, Amsterdam, the Netherlands; Simon Gilbody,

University of York, Heslington, York, UK; John P.A. Ioannidis, Stanford University, Stanford, California, USA; Lorie A. Kloda, Concordia University, Montréal, Québec, Canada; Dean McMillan, University of York, Heslington, York, UK; Scott B. Patten, University of Calgary, Calgary, Alberta, Canada; Ian Shrier, McGill University, Montréal, Québec, Canada; Roy C. Ziegelstein, Johns Hopkins University School of Medicine, Baltimore, Maryland, USA; Dickens H. Akena, Makerere University College of Health Sciences, Kampala, Uganda; Bruce Arroll, University of Auckland, Auckland, New Zealand; Liat Ayalon, Bar Ilan University, Ramat Gan, Israel; Hamid R. Baradaran, Iran University of Medical Sciences, Tehran, Iran; Murray Baron, McGill University, Montréal, Québec, Canada; Anna Beraldi, Lehrkrankenhaus der Technischen Universität München, Munich, Germany; Charles H. Bombardier, University of Washington, Seattle, Washington, USA; Peter Butterworth, The University of Melbourne, Melbourne, Australia; Gregory Carter, University of Newcastle, New South Wales, Australia; Marcos H. Chagas, University of São Paulo, Ribeirão Preto, Brazil; Juliana C. N. Chan, The Chinese University of Hong Kong, Hong Kong Special Administrative Region, China; Rushina Cholera, University of North Carolina at Chapel Hill School of Medicine, Chapel Hill, North Carolina, USA; Neerja Chowdhary, Clinical practice, Mumbai, India; Kerrie Clover, University of Newcastle, New South Wales, Australia; Yeates Conwell, University of Rochester Medical Center, Rochester, New York, USA; Janneke M. de Man-van Ginkel, University Medical Center Utrecht, Utrecht, The Netherlands; Jaime Delgadillo, University of Sheffield, Sheffield, UK; Jesse R. Fann, University of Washington, Seattle, Washington, USA; Felix H. Fischer, Charité - Universitätsmedizin Berlin, Berlin, Germany; Benjamin Fischler, Private Practice, Brussels, Belgium; Daniel Fung, Duke-NUS Medical School, Singapore; Bizu Gelaye, Harvard T. H. Chan School of Public Health, Boston, Massachusetts, USA; Felicity Goodyear-Smith, University of Auckland, Auckland, New Zealand; Catherine G. Greeno, University of Pittsburgh, Pittsburgh, Pennsylvania, USA; Brian J. Hall, University of Macau, Macau Special Administrative Region, China; Patricia A. Harrison, City of Minneapolis Health Department, Minneapolis, Minnesota, USA; Martin Harter, University Medical Center Hamburg-Eppendorf, Hamburg, Germany; Ulrich Hegerl, German Depression Foundation, Leipzig, Germany; Leanne Hides, University of Queensland, Brisbane, Queensland, Australia; Stevan E. Hobfoll, STAR-Stress, Anxiety & Resilience Consultants, Chicago, Illinois, USA; Marie Hudson, McGill University, Montréal, Québec, Canada; Thomas Hyphantis, University of Ioannina, Ioannina, Greece; Masatoshi Inagaki, Shimane University, Shimane, Japan; Khalida Ismail, King's College London Weston Education Centre, London, UK; Nathalie Jetté, Ichan School of Medicine at Mount Sinai, New York, New York, USA; Mohammad E. Khamseh, Iran University of Medical Sciences, Tehran, Iran; Kim M. Kiely, University of New South Wales, Sydney, Australia; Yunxin Kwan, Tan Tock Seng Hospital, Singapore; Femke Lamers, Amsterdam UMC, Amsterdam, The Netherlands; Shen-Ing Liu, Mackay Memorial Hospital, Taipei, Taiwan; Manote Lotrakul, Mahidol University, Bangkok, Thailand; Sonia R. Loureiro, University of São Paulo, Ribeirão Preto, Brazil; Bernd Löwe, University Medical Center Hamburg-Eppendorf, Hamburg, Germany; Laura Marsh, Baylor College of Medicine, Houston and Michael E. DeBakey Veterans Affairs Medical Center, Houston, Texas, USA; Anthony McGuire, St. Joseph's College, Standish, Maine, USA; Sherina Mohd Sidik, Universiti Putra Malaysia, Serdang, Selangor, Malaysia; Tiago N. Munhoz, Federal University of Pelotas, Pelotas, Brazil; Kumiko Muramatsu, Graduate School of Niigata Seiryo University, Niigata, Japan; Flávia L. Osório, University of São Paulo, Ribeirão Preto, Brazil; Vikram Patel, Harvard Medical School, Boston, Massachusetts, USA; Brian W. Pence, The University of North Carolina at Chapel Hill,

Chapel Hill, North Carolina, USA; Philippe Persoons, Katholieke Universiteit Leuven, Leuven, Belgium; Angelo Picardi, Italian National Institute of Health, Rome, Italy; Katrin Reuter, Group Practice for Psychotherapy and Psycho-oncology, Freiburg, Germany; Alasdair G. Rooney, University of Edinburgh, Edinburgh, Scotland, UK; Iná S. Santos, Federal University of Pelotas, Pelotas, Brazil; Juwita Shaaban, Universiti Sains Malaysia, Kelantan, Malaysia; Abbey Sidebottom, Allina Health, Minneapolis, Minnesota, USA; Adam Simning, University of Rochester Medical Center, Rochester, New York; Lesley Stafford, Royal Women's Hospital, Parkville, Australia; Sharon C. Sung, Duke-NUS Medical School, Singapore; Pei Lin Lynnette Tan, Tan Tock Seng Hospital, Singapore; Alyna Turner, University of Newcastle, New South Wales, Newcastle, Australia; Christina M. van der Feltz-Cornelis, University of York, York, UK; Henk C. van Weert, Amsterdam University Medical Centers, Location AMC, Amsterdam, the Netherlands; Paul A. Vöhringer, Universidad de Chile, Santiago, Chile; Jennifer White, Monash University, Melbourne, Australia; Mary A. Whooley, Veterans Affairs Medical Center, San Francisco, California, USA; Kirsty Winkley, King's College London, Waterloo Road, London, UK ; Mitsuhiko Yamada, National Center of Neurology and Psychiatry, Tokyo, Japan; Yuying Zhang, The Chinese University of Hong Kong, Hong Kong Special Administrative Region, China.

**Corresponding Author:**

Andrea Benedetti, Research Institute of the McGill University Health Center, 3D.59, 5252 boulevard de Maisonneuve, Montreal, Quebec, Canada

Email: andrea.benedetti@mcgill.ca

**Abstract**

Researchers increasingly use meta-analysis to synthesize the results of several studies in order to estimate a common effect. When the outcome variable is continuous, standard meta-analytic approaches assume that the primary studies report the sample mean and standard deviation of the outcome. However, when the outcome is skewed, authors sometimes summarize the data by reporting the sample median and one or both of (i) the minimum and maximum values and (ii) the first and third quartiles, but do not report the mean or standard deviation. To include these studies in meta-analysis, several methods have been developed to estimate the sample mean and standard deviation from the reported summary data. A major limitation of these widely used methods is that they assume that the outcome distribution is normal, which is unlikely to be tenable for studies reporting medians. We propose two novel approaches to estimate the sample mean and standard deviation when data are suspected to be non-normal. Our simulation results and empirical assessments show that the proposed methods often perform better than the existing methods when applied to non-normal data.

**Keywords:** meta-analysis, median, first quartile, third quartile, minimum value, maximum value

**Introduction**

Meta-analysis is a statistical approach for pooling data from related studies that is widely used to provide evidence for medical research. To pool studies in an aggregate data meta-analysis, each study must contribute an effect measure (e.g., the sample mean for one-group studies, the sample means for two-group studies) of the outcome and its variance. However, primary studies may differ in the effect measures reported. Although the sample mean is the usual effect measure reported for continuous outcomes, authors often report the sample median when data are skewed and may not report the mean.[1] This occurs commonly for time-based outcomes, such as time delays in the diagnosis and treatment of tuberculosis[2,3] or colorectal cancer[4] or length of hospital stay[5-7]. Other examples in medical research include muscle strength and mass[8], molecular concentration levels[9], tumor sizes[10], motor impairment scores[11], and intraoperative blood loss[12]. When primary studies report the sample median of an outcome, they typically report the sample size and one or both of (i) the sample minimum and maximum values and (ii) the first and third quartiles.

The same effect measure must be obtained from all primary studies in an aggregate data meta-analysis. In order to meta-analyze a collection of studies in which some report the sample mean and others report the sample median, Hozo et al.[13], Bland[14], Wan et al.[15], Kwon and Reis[16], and Luo et al.[17] have recently published methods to estimate the sample mean and standard deviation from studies that report medians. These methods have been widely used to meta-analyze the means for one-group studies and the raw or standardized difference of means for two-group studies. Reflecting how commonly these methods are used, Google Scholar listed 2,871 articles citing Hozo et al.[13] and 601 articles citing Wan et al.[15] as of March 12, 2019.

Commonly used methods that have been proposed to estimate the sample mean and standard deviation in this context can be divided into formula-based methods and simulation-based methods. The methods developed by Luo et al.[17] and Wan et al.[15] are the best-performing formula-based methods for estimating the sample mean and standard deviation, respectively. A major limitation of these methods is that they assume the outcome variable is normally distributed, which may be unlikely because otherwise the authors would have reported the mean. Consequently, Kwon and Reis[16] recently proposed a simulation-based method which is based on different parametric assumptions of the outcome variable. Although the Kwon and Reis[16] sample mean estimator has not been compared to the formula-based method of Luo et al.[17], their proposed standard deviation estimator performed better than the formula-based method of Wan et al.[15] for skewed data when the assumed parametric family is correct. Two limitations of this simulation-based method are that it is computationally expensive and requires users to write their own distribution-specific code.

We propose two novel methods to estimate the sample mean and standard deviation for skewed data when the underlying distribution is unknown. The proposed methods overcome several limitations of the existing methods, and we demonstrate that the proposed approaches often perform better than the existing methods when applied to skewed data.

The objectives of this paper are to describe the existing and proposed methods for estimating the sample mean and standard deviation, systematically evaluate their performance in a simulation study, and empirically evaluate their performance on real-life data sets.

In the following section, we describe the existing and proposed methods. In 'Results', we report the results of a simulation investigating the performance of the methods. We illustrate these methods on an example data set and evaluate their accuracy in 'Example'. In 'Discussion', we summarize our findings and provide recommendations for data analysts.

**Methods**

Throughout this paper, we use the following notation for sample summary statistics: minimum value ($Q_{\min}$), first quartile ($Q_1$), median ($Q_2$), third quartile ($Q_3$), maximum value ($Q_{\max}$), mean ($\bar{x}$), standard deviation ($s_x$), and sample size ($n$). As investigated in previous studies[13-17], we consider the following sets of summary statistics that may be reported by a study, denoted by Scenario 1 ($S_1$), Scenario 2 ($S_2$), and Scenario 3 ($S_3$):

$$S_1 = \{Q_{\min}, Q_2, Q_{\max}, n\}$$
$$S_2 = \{Q_1, Q_2, Q_3, n\}$$
$$S_3 = \{Q_{\min}, Q_1, Q_2, Q_3, Q_{\max}, n\}.$$

**Existing Methods**

*Formula-based Methods: Luo et al.[17] and Wan et al.[15]*

The sample mean estimator of Luo et al.[17] and the sample standard deviation estimator of Wan et al.[15] are formula-based methods that are derived from the assumption that the outcome variable is normally distributed.

Luo et al. developed the following sample mean estimators in scenarios $S_1$, $S_2$, and $S_3$:

$$\bar{x} = \left(\frac{4}{4+n^{0.75}}\right)\frac{Q_{min}+Q_{max}}{2} + \left(\frac{n^{0.75}}{4+n^{0.75}}\right)Q_2 \qquad \text{in } S_1$$

$$\bar{x} = \left(0.7 + \frac{0.39}{n}\right)\frac{Q_1+Q_3}{2} + \left(0.3 - \frac{0.39}{n}\right)Q_2 \qquad \text{in } S_2$$

$$\bar{x} = \left(\frac{2.2}{2.2+n^{0.75}}\right)\frac{Q_{min}+Q_{max}}{2} + \left(0.7 - \frac{0.72}{n^{0.55}}\right)\frac{Q_1+Q_3}{2}$$
$$+ \left(0.3 + \frac{0.72}{n^{0.55}} - \frac{2.2}{2.2+n^{0.75}}\right)Q_2 \qquad \text{in } S_3$$

Building on the sample mean estimators of Hozo et al.[13], Wan et al.[15], and Bland[14] in $S_1$, $S_2$, and $S_3$, respectively, this method optimally weights the median (in $S_1$, $S_2$, and $S_3$), the average of the minimum and maximum values (in $S_1$ and $S_3$), and the average of the first and third quartiles (in $S_2$ and $S_3$). The weights are set to minimize the mean squared error of the estimator. Numerical simulations have demonstrated that the method of Luo et al. has considerably lower relative mean squared error (RMSE) compared to the method of Bland in $S_3$ and has comparable RMSE to the method Wan et al. in $S_2$ under normal and skewed distributions.

Wan et al. proposed the following sample standard deviation estimators in scenarios $S_1$, $S_2$, and $S_3$:

$$s_x = \frac{Q_{\max} - Q_{\min}}{2\Phi^{-1}\left(\frac{n - 0.375}{n + 0.25}\right)} \quad \text{in } S_1$$

$$s_x = \frac{Q_3 - Q_1}{2\Phi^{-1}\left(\frac{0.75n - 0.125}{n + 0.25}\right)} \quad \text{in } S_2$$

$$s_x = \frac{Q_{\max} - Q_{\min}}{4\Phi^{-1}\left(\frac{n - 0.375}{n + 0.25}\right)} + \frac{Q_3 - Q_1}{4\Phi^{-1}\left(\frac{0.75n - 0.125}{n + 0.25}\right)} \quad \text{in } S_3$$

The standard deviation estimators of Wan et al. are derived using relationships between the distribution standard deviation and the expected values of order statistics for normally distributed data. The expected values of the minimum and maximum values and first and third quartiles are estimated by the respective sample values. The expected value of other order statistics are estimated using Blom's method[18].

Wan et al. were the first to propose a standard deviation estimator in $S_2$. Wan et al. showed that their estimator in $S_1$ and $S_3$ outperformed the previously developed sample standard deviation estimators of Hozo et al.[13] and Bland[14], respectively, in regards to average relative error.

For the purpose of the analyses presented herein, we refer to the approach which uses the method of Luo et al. to estimate the sample mean and the method of Wan et al. to estimate the sample standard deviation as the Luo/Wan method.

*Simulation-based Method: Kwon and Reis[16]*

Kwon and Reis[16] proposed a method based on applying approximate Bayesian computation (ABC) to estimate the sample mean and standard deviation in scenarios $S_1$, $S_2$, and $S_3$. Unlike the methods of Luo et al. and Wan et al. which assume that the outcome variable is normally distributed, this method can be applied under different parametric assumptions of the underlying distribution (i.e., normal and skewed distributions). Throughout this paper, we will refer to the approach of Kwon and Reis[16] as the ABC method.

The ABC method can be briefly described as follows. In the context where the underlying distribution is unknown a priori, the several candidate parametric families of distributions are specified, namely the normal, log-normal, exponential, beta, and Weibull distributions. The parameters of each distribution are estimated by applying the ABC rejection sampling algorithm (described below) proposed by Kwon and Reis[19]. This version of the algorithm, given in Kwon and Reis[19], builds on that of Kwon and Reis[16] to incorporate several candidate parametric families of distributions in a more computationally efficient manner.

In brief, the ABC rejection sampling algorithm samples parameter values of the candidate distributions and simulates pseudo data (i.e., sample median and one or both of (i) the minimum and maximum values and (ii) the first and third quartiles). If the pseudo data are sufficiently close to the summary data reported by a study, the parameter values are accepted. For each candidate distribution, the distributions of the accepted parameters approximate their respective posterior distributions after a large number of iterations of the algorithm. The candidate distribution with the highest marginal posterior probability is selected. The means of the respective posterior distributions are used to estimate the parameters of the selected distribution.

Kwon and Reis[16] demonstrated that, provided the candidate distribution is correctly specified and the sample size is sufficiently large (e.g., $n \geq 100$), their proposed ABC method outperformed the sample mean estimators of Hozo et al.[13] (for $S_1$), Wan et al.[15] (for $S_2$), and Bland[14] (for $S_3$) and outperformed the standard deviation estimators of Wan et al. for skewed distributions.

**Proposed Methods**

The following two subsections describe the proposed methods for estimating the sample mean and standard deviation from $S_1$, $S_2$, and $S_3$ summary measures. The R package 'estmeansd' available on CRAN implements both of the proposed methods.[20] Additionally, the webpage https://smcgrath.shinyapps.io/estmeansd/ provides a graphical user interface for using these methods.

**Quantile Estimation (QE) Method**

The QE method was originally introduced in McGrath et al.[21] for estimating the variance of the median when summary measures of $S_1$, $S_2$, or $S_3$ are provided. Here, we describe how the QE method can be applied to estimate the sample mean and standard deviation in these contexts.

We pre-specify several candidate parametric families of distributions for the outcome variable, namely the normal, log-normal, gamma, beta, and Weibull. The parameters of each candidate distribution are estimated by minimizing the distance between the observed and distribution quantiles. Let $F_\theta^{-1}$ denote the quantile function of a given candidate distribution parameterized by $\theta$. Then, the objective function corresponding to the distribution, denoted by $S(\theta)$, is given by

$$S(\theta) = \left(F_\theta^{-1}(1/n) - Q_{\min}\right)^2 + \left(F_\theta^{-1}(0.5) - Q_2\right)^2 + \left(F_\theta^{-1}(1 - 1/n) - Q_{\max}\right)^2 \quad \text{in } S_1$$

$$S(\theta) = \left(F_\theta^{-1}(0.25) - Q_1\right)^2 + \left(F_\theta^{-1}(0.5) - Q_2\right)^2 + \left(F_\theta^{-1}(0.75) - Q_3\right)^2 \quad \text{in } S_2$$

$$S(\theta) = \left(F_\theta^{-1}(1/n) - Q_{\min}\right)^2 + \left(F_\theta^{-1}(0.25) - Q_1\right)^2 + \left(F_\theta^{-1}(0.5) - Q_2\right)^2$$
$$+ \left(F_\theta^{-1}(0.75) - Q_3\right)^2 + \left(F_\theta^{-1}(1 - 1/n) - Q_{\max}\right)^2 \quad \text{in } S_3$$

Details concerning the implementation of the optimization algorithm for minimizing $S(\theta)$ are provided in Appendix A.

The distribution with the best fit (i.e., yielding the smallest value of $S(\hat{\theta})$ where $\hat{\theta}$ denotes the estimated parameters of the given distribution) is assumed to be the underlying distribution of the sample. The sample mean and standard deviation are estimated by the mean and standard deviation of the selected distribution.

**Box-Cox (BC) Method**

Luo et al.[17] and Wan et al.[15] assumed that a sample $x$ of interest follows a normal distribution. To make this assumption more tenable for skewed data, we incorporate Box-Cox transformations into the methods of Luo et al. and Wan et al. The proposed method, which we denote by BC, applies Box-Cox transformations to the quantiles of $x$ and assumes that the underlying distribution of the transformed data is normal.

In brief, the BC method consists of the following four steps. First, an optimization algorithm, such as the algorithm of Brent[22], optimizes the power parameter $\lambda$ such that distribution of the transformed data is most likely to be normal. Letting $f_\lambda$ denote the Box-Cox transformation, the quantiles of $x$ are transformed into the quantiles of $f_\lambda(x)$. Afterwards, the methods of Luo et al. and Wan et al. are applied to estimate the mean and standard deviation of $f_\lambda(x)$, respectively. Finally, the mean and standard deviation of $f_\lambda(x)$ are inverse-transformed into the mean and standard deviation of $x$.

Box-Cox transformations $f_\lambda$ are defined as follows:

$$f_\lambda(x_i) = y_i = \begin{cases} \dfrac{x_i^\lambda - 1}{\lambda} & \text{if } \lambda \neq 0 \\ \ln(x_i) & \text{if } \lambda = 0 \end{cases}$$

Equivalently, inverse Box-Cox transformations $f_\lambda^{-1}$ are defined as follows:

$$f_\lambda^{-1}(y_i) = x_i = \begin{cases} (\lambda \cdot y_i + 1)^{1/\lambda} & \text{if } \lambda \neq 0 \\ \exp(y_i) & \text{if } \lambda = 0 \end{cases}$$

Box and Cox[23] argued that Box-Cox transformations can transform a dataset into a more normally-distributed dataset. Moreover, for every value of $\lambda$, $f_\lambda$ is monotonically increasing. Therefore, any i[th] order statistic of an untransformed dataset, after transformation, is still the i[th] order statistic of the corresponding transformed dataset, and vice versa.

The optimization step for finding $\lambda$ can be described as follows. In $S_1$ and $S_2$, $\lambda$ is chosen so that the transformed minimum and maximum values (in $S_1$) or first and third quartiles (in $S_2$) are equidistant from the median, making the transformed data to be most likely symmetric and therefore most normally distributed. Specifically, the BC method finds a finite value of $\lambda$ such that

$$f_\lambda(Q_{\max}) - f_\lambda(Q_2) = f_\lambda(Q_2) - f_\lambda(Q_{\min})$$

in $S_1$ and

$$f_\lambda(Q_3) - f_\lambda(Q_2) = f_\lambda(Q_2) - f_\lambda(Q_1)$$

in $S_2$. In $S_3$, a value of $\lambda$ cannot necessarily be found such that both the first and third quartiles as well as the minimum and maximum values are equidistant from the median. Therefore, $\lambda$ is found by

$$\operatorname*{argmin}_{\lambda} \left[ \left(f_\lambda(Q_3) - f_\lambda(Q_2) - (f_\lambda(Q_2) - f_\lambda(Q_1))\right)^2 \right.$$
$$\left. + \left(f_\lambda(Q_{\max}) - f_\lambda(Q_2) - (f_\lambda(Q_2) - f_\lambda(Q_{\min}))\right)^2 \right]$$

Appendix B describes the implementation of the optimization algorithm used to find $\lambda$.

Then, the BC method applies the Box-Cox transformations with this value of $\lambda$ on the quantiles of $x$. That is, the BC method transforms $\{Q_{\min}, Q_2, Q_{\max}\}$ into $\{f_\lambda(Q_{\min}), f_\lambda(Q_2), f_\lambda(Q_{\max})\}$ in $S_1$, $\{Q_1, Q_2, Q_3\}$ into $\{f_\lambda(Q_1), f_\lambda(Q_2), f_\lambda(Q_3)\}$ in $S_2$, and $\{Q_{\min}, Q_1, Q_2, Q_3, Q_{\max}\}$ into $\{f_\lambda(Q_{\min}), f_\lambda(Q_1), f_\lambda(Q_2), f_\lambda(Q_3), f_\lambda(Q_{\max})\}$ in $S_3$.

Let $N'(\mu, \sigma^2) \sim N(\mu, \sigma^2)$ conditional on $N'(\mu, \sigma^2) \in [f(0), 2\mu - f(0)]$. Equivalently, $N'(\mu, \sigma^2)$ is the symmetrically truncated $N(\mu, \sigma^2)$ bounded within the support $[f(0), 2\mu - f(0)]$. Then, the BC method assumes that $f_\lambda(x) \sim N'(\mu, \sigma^2)$ for some $\mu$ and $\sigma$ and uses the methods of Luo et al. and Wan et al. to calculate $\mu$ and $\sigma$, respectively. Finally, the assumption made by the BC method implies that $x \sim f_\lambda^{-1}(N'(\mu, \sigma^2))$. Therefore, the mean and standard deviation of $f_\lambda^{-1}(N'(\mu, \sigma^2))$ are approximately $\bar{x}$ and $s_x$.

The mean and standard deviation of $f_\lambda^{-1}(N'(\mu, \sigma^2))$ are found as follows. Let $\phi$ and $\Phi$ be the probability density function and cumulative distribution function of the standard normal distribution, respectively. The following two equations describe the mean and variance of $f_\lambda^{-1}(N'(\mu, \sigma^2))$, respectively:

$$E[f_\lambda^{-1}(N'(\mu, \sigma^2))] = \int_{x=f_\lambda(0)}^{x=2\mu-f_\lambda(0)} \phi\left(\frac{x-\mu}{\sigma}\right) \frac{f_\lambda^{-1}(x)}{\sigma(\Phi(\mu) - \Phi(-\mu))} \partial x$$

$$\mathrm{Var}[f_\lambda^{-1}(N'(\mu, \sigma^2))] = \int_{x=f_\lambda(0)}^{x=2\mu-f_\lambda(0)} \phi\left(\frac{x-\mu}{\sigma}\right) \frac{\left(f_\lambda^{-1}(x) - E[f_\lambda^{-1}(N'(\mu, \sigma^2))]\right)^2}{\sigma(\Phi(\mu) - \Phi(-\mu))} \partial x$$

Numerical integration can solve the two above equations. Moreover, the following Monte-Carlo simulation can compute the mean and standard deviation of $f_\lambda^{-1}(N'(\mu, \sigma^2))$: first, generate an independent and identically distributed random sample $R$ from $N(\mu, \sigma^2)$; next, let the new $R$ be $\{r \in R : r \in [f(0), 2\mu - f(0)]\}$, or equivalently, remove any value in $R$ that is not within the range $[f(0), 2\mu - f(0)]$; then, calculate the sample mean and sample standard deviation of $R$; finally, the sample mean and sample standard deviation are estimated as the mean and standard deviation of $f_\lambda^{-1}(N'(\mu, \sigma^2))$. The application of the BC method in this work uses Monte-Carlo simulation to compute the mean and standard deviation of $f_\lambda^{-1}(N'(\mu, \sigma^2))$.

Recall that $N'(\mu, \sigma^2)$ is the symmetrically truncated $N(\mu, \sigma^2)$ with support $[f(0), 2\mu - f(0)]$. In fact, $N'(\mu, \sigma^2) \sim f_{\lambda=1}^{-1}(N'(\mu, \sigma^2))$, and $LN(\mu, \sigma^2) \sim f_{\lambda=0}^{-1}(N'(\mu, \sigma^2))$. Therefore, both the normal distribution truncated within the support $[f(0), 2\mu - f(0)]$ and log-normal distribution are special cases of $f_\lambda^{-1}(N'(\mu, \sigma^2))$.

**Design of Simulation Study**

We conducted a simulation study to systematically compare the performance of the existing and proposed approaches when the truth is known.

To be consistent with the work already conducted in this area, we generated data from the same distributions considered in previous studies[13-17]. As used by Bland[14], we used the normal distribution with $\mu = 5$ and $\sigma = 1$, the log-normal distribution with $\mu = 5$ and $\sigma = 0.25$, the log-normal distribution with $\mu = 5$ and $\sigma = 0.5$, and the log-normal distribution $\mu = 5$ and $\sigma = 1$ in our primary analyses to investigate the effect of skewness on the performance of the sample mean and standard deviation estimators. In sensitivity analyses, we considered the following distributions used in several other studies[13, 15-17]: the normal distribution with $\mu = 50$ and $\sigma = 17$, the log-normal distribution with $\mu = 4$ and $\sigma = 0.3$, the exponential distribution with $\lambda = 10$, the beta distribution with $\alpha = 9$ and $\beta = 4$, and the Weibull distribution with $\lambda = 2$ and $k = 35$.

For each distribution, a sample of size $n$ was drawn to simulate data from a primary study. Then, the appropriate summary statistics (i.e., $S_1$, $S_2$, or $S_3$) were calculated from this sample. The Luo/Wan, ABC, QE, and BC methods were each applied to the summary data in order to estimate the sample mean and standard deviation. We will refer to these estimates as the "derived estimated sample means and standard deviations". The true sample mean and standard deviation were then compared to the derived estimated sample means and standard deviations. As used in previous studies[13, 15, 16], the relative error was used as the performance measure. The relative error is defined by

$$\text{relative error of } x = \frac{\text{estimated } x - \text{true } x}{\text{true } x}.$$

We used the following sample sizes in our simulations: 25, 50, 75, 100, 150, 200, 250, 300, 350, 400, 450, 500, 550, 600, 650, 700, 750, 800, 850, 900, 950, 1 000. A total of 1 000 repetitions were performed for each combination of data generation parameters under scenarios $S_1$, $S_2$, and $S_3$. The average relative error (ARE) was calculated over the 1 000 repetitions for each combination of data generation parameters.

**Results of Simulation Study**

In the following subsections, we present the results of the simulation study using the set of outcome distributions considered by Bland[14], as these distributions were selected to investigate the effect of skewness on the estimators. The results of the sensitivity analyses where we used the set of outcome distribution used by other authors[13, 15-17] is given in Section 1 of Supplementary Material.

Because the simulation results in scenarios $S_1$ and $S_3$ were similar, the $S_3$ simulation results are presented in Section 2 of Supplementary Material for parsimony. Additionally, as the focus of this paper is on the analysis of non-normal data, all simulation results where data were generated from a normal distribution are presented in Section 3 of Supplementary Material.

**Comparison of Methods Under Scenario $S_1$**

Figure 1 displays the ARE of all sample mean and standard deviation estimators under scenario $S_1$. As the skewness (i.e., the $\sigma$ parameter) of the log-normal distribution increased, the magnitude of the AREs generally increased for the sample mean and standard deviation estimators, but was inconsequential for the BC method. Moreover, all methods had considerably larger AREs for estimating the sample standard deviation compared to estimating the sample mean.

For estimating the sample mean, the BC method performed best under each distribution and nearly all sample sizes ($n$) considered in Figure 1; the BC method was nearly unbiased, yielding AREs of magnitude less than 0.004, 0.008, and 0.020 in the Log-Normal(5,0.25), Log-Normal(5,0.5), and Log-Normal(5,1), cases, respectively. Contrary to the Luo et al. and ABC sample mean estimators which became more biased as $n$ increased (e.g., ARE $= -0.22$ for Luo et al. and ARE $= -0.40$ for ABC when $n = 1\,000$ in Log-Normal(5,1)), the performance of the QE sample mean estimator improved as $n$ increased. The QE sample mean estimator became preferred over the Luo et al. and ABC sample mean estimators when $n \geq 300$. However, the QE method always performed worse than the BC method in regards to ARE in Figure 1.

The BC method performed best for estimating the sample standard deviation, achieving AREs of magnitude less than 0.03 in nearly all scenarios investigated in Figure 1. Although the QE standard deviation estimator performed better as $n$ increased, this method typically resulted in larger AREs compared to the ABC and BC methods. Additionally, the QE and ABC standard deviation estimators often yielded large ARE values when sample sizes were small (i.e., $n \leq 75$), especially for skewed outcomes.

Model selection highly differed between the QE and ABC methods when the outcome distribution was Log-Normal(5,0.25). For this outcome distribution, the percentage of repetitions where the ABC method selected the log-normal distribution ranged between 0.6% (when $n = 75$) and 5.3% (when $n = 900$). In all repetitions where the log-normal distribution was not selected, the ABC method selected the normal distribution. The QE method, on the other hand, selected the log-normal distribution between 58.1% (when $n = 25$) to 82.3% (when $n = 1\,000$) of repetitions. Moreover, the QE method had comparable performance in the repetitions where it did not select the log-normal distribution (e.g., AREs ranging between -0.01 and 0.01 for estimating the sample mean and between 0.07 and 0.11 for estimating the standard deviation in these repetitions). Model selection improved for the QE and ABC methods as $n$ and the skewness of the log-normal distribution increased. For example, in the Log-Normal(5,1) case, the ABC selected the log-normal distribution in at least 99.9% of the repetitions for all $n$ and the QE method selected the log-normal distribution in at least 99% of the repetitions for all $n \geq 50$.

**Comparison of Methods Under Scenario $S_2$**

Figure 2 gives the ARE of all methods under scenario $S_2$. As in scenario $S_1$, we found that (i) the skewness of the underlying distribution strongly affected the performance of the sample mean and standard deviation estimators, and (ii) the sample mean estimators typically had AREs with smaller magnitude.

The BC and QE sample mean estimators performed comparably to each other in most scenarios investigated in Figure 2. In the Log-Normal(5,0.25) case, these two methods performed best. In the Log-Normal(5,0.5) and Log-Normal(5,1) cases, the BC, QE, and ABC methods all performed comparably to each other and the Wan et al. method performed considerably worse. Additionally, for small $n$ and skewed data, the ABC sample mean estimator gave highly biased estimates (e.g., ARE $= 0.59$ when $n = 25$ in Log-Normal(5,1)).

Similar trends held for the corresponding sample standard deviation estimators. The QE and BC methods performed best in the Log-Normal(5,0.25) case, and the ABC, QE, and BC methods performed best and comparably in the Log-Normal(5,0.5) and Log-Normal(5,1) cases. Moreover, for small sample sizes in the Log-Normal(5,1) case, the ABC method yielded very large ARE values (e.g., ARE $= 3.48$ when $n = 25$ in Log-Normal(5,1)).

Lastly, model selection performance was similar to that observed in $S_1$. ABC model selection performed poorly in the Log-Normal(5,0.25) case, as it selected the normal distribution for all 1 000 repetitions under all values of $n$. The QE method, on the other hand, selected the log-normal distribution in the majority of repetitions under all values of $n$. The performance of the QE method slightly worsened in repetitions where the log-normal solution was not selected (e.g.,

AREs ranging between -0.02 to -0.01 for estimating the sample mean and between -0.08 and -0.03 for estimating the sample standard deviation in these repetitions) As $n$ and the skewness of the underlying log-normal distribution increased, the log-normal distribution was increasingly selected by the ABC and QE methods. For instance, in the Log-Normal(5,1) case, the ABC method selected the log-normal distribution in at least 96% of the repetitions under all $n$ and the QE method selected the log-normal distribution in at least 90% of the repetitions for all $n \geq 250$.

**Example**

In this section, we illustrate the use of the existing and proposed methods when applied to a real-life meta-analysis of a continuous, skewed outcome. Specifically, we used data collected for an individual participant data (IPD) meta-analysis of the diagnostic accuracy of the Patient Health Questionnaire-9 (PHQ-9) depression screening tool.[24, 25] We chose to use data from an IPD meta-analysis because 1) $S_1$, $S_2$, and $S_3$ summary data can be obtained from each study and 2) the true study-specific sample means and standard deviations are available.

Our analysis focused on the patient scores of the PHQ-9, which is a self-administered screening tool for depression. PHQ-9 scores are measured on a scale from 0 to 27, where higher scores are indicative of higher depressive symptoms. Previous studies have found that the distribution of PHQ-9 scores in the general population is right-skewed[26-28].

For each of the 58 primary studies, we calculated the sample median, minimum and maximum values, and first and third quartiles of the PHQ-9 scores of all patients in order to mimic the scenarios where an aggregate data meta-analysis extracts $S_1$, $S_2$, or $S_3$ summary data. Then, we applied the existing and proposed methods to this summary data to estimate study-specific sample means and standard deviations – we refer to these as the "derived estimated sample means and standard deviations". Section 4 of Supplementary Material presents the study-specific $S_3$ summary data.

Some primary studies used weighted sampling. When extracting $S_1$, $S_2$, and $S_3$ summary data from these studies, weighted sample quantiles were used.[29] Additionally, weighted sample means and standard deviations were used as the true values for the sample mean and standard deviation, respectively, for studies with weighted sampling.

As PHQ-9 scores are integer-valued, PHQ-9 scores of 0 were observed in most of the primary studies. However, a minimum value and/or first quartile value of 0 result in complications for the QE and ABC methods when estimating the parameters of the log-normal distribution, as the prior bounds for the ABC method and the parameter constraints for the QE method implicitly assume that the extracted summary data are strictly positive. Therefore, when applying all methods, a value of 0.5 was added to the extracted summary data. After estimating the sample mean and standard deviation from the shifted summary data, 0.5 was subtracted from the estimated sample mean.

We compared the derived estimated sample means and standard deviations to the true sample means and standard deviations (Table 1). The QE and BC methods were considerably less biased than the existing methods for estimating the sample mean under $S_1$, $S_2$, and $S_3$. The QE sample mean estimator performed best under $S_1$ and the BC sample mean estimator performed best under $S_2$ and $S_3$. Trends were less conclusive for estimating the standard deviation. The QE method standard deviation estimator was the least biased under $S_1$ and $S_3$ and the standard deviation estimator of Wan et al. was the least biased under $S_2$. The high ARE value of the ABC method for estimating the standard deviation under $S_2$ was due to very large relative error values (relative error $> 10$) when applied to the Osorio et al. 2009, Ayalon et al. 2010, and Twist et al. 2013 studies.

We meta-analyzed the PHQ-9 scores using the true study-specific sample means and standard deviations (Figure 3) and compared this to a meta-analysis using the derived estimated study-specific sample means and standard deviations (Table 2). The restricted maximum likelihood method was used to estimate heterogeneity in all meta-analyses.[30] The QE and BC methods were less biased for estimating the pooled mean compared to the existing methods in $S_1$, $S_2$, and $S_3$. The QE method had relative error closest to zero for estimating the pooled mean in $S_1$ and $S_3$ and the BC method had relative error closest to zero in $S_2$. As one may expect, QE and BC methods performed best in $S_3$ for estimating the pooled mean, yielding relative errors of -0.0054 and 0.0074, respectively.

The primary studies were highly heterogeneous. When using the true study-specific sample means and standard deviations, the $I^2 = 98.15\%$.[31] The Luo/Wan, ABC, QE, and BC methods yielded similar estimates of $I^2$; using 98.15% as the true value of $I^2$, all four methods had relative errors between $-0.02$ and $0.02$ for estimating $I^2$ in $S_1$, $S_2$, and $S_3$.

Lastly, we investigated the skewness of the PHQ-9 scores. To mimic how data analysts may evaluate skewness based on available summary data, we used Bowley's coefficient to quantify skewness, as it only depends on $S_2$ summary data.[32] Bowley's coefficient values range from -1 to 1, where positive values indicate right skew and negative values indicate left skew. The average value of Bowley's coefficient taken over all 58 primary studies was 0.18, indicating moderate right skewness. Moreover, the ABC and QE methods suggested non-normality in many of the primary studies. When given $S_2$ data, the ABC method selected the normal distribution for 50% of studies and the log-normal for the other 50% of studies. The QE method selected the normal distribution for 21% of studies, the log-normal for 22% of studies, the gamma for 26% of studies, and the Weibull for 31% of studies.

We performed additional analyses to explore the sensitivity of the addition of 0.5 to all summary data. When adding 0.1 or 0.01 to all summary data, similar results for the Luo/Wan, QE, and BC methods were obtained. However, the performance of the ABC method considerably worsened for smaller values added to the summary data, especially in $S_2$. For instance, the ABC method had ARE of 0.60 for estimating the sample mean and 11.15 for estimating the sample standard deviation in $S_2$ when 0.01 was added to all summary data.

**Discussion**

We proposed two methods to estimate the sample mean and standard deviation from commonly reported quantiles in meta-analysis. Because studies typically report the sample median and other sample quantiles when data are skewed, our analyses focused on the application of the proposed

QE and BC methods to skewed data. We compared the QE and BC methods to the widely used methods of Wan et al.[15], Luo et al.[17], and Kwon and Reis[16] in a simulation study and in a real-life meta-analysis.

We found that the QE and BC sample mean estimators performed well, typically yielding average relative error values approaching zero as the sample size increased. In the simulation study, the QE and BC sample mean estimators performed better than the methods of Luo et al. in nearly all scenarios and often performed better than the ABC method of Kwon and Reis[16]. In our empirical evaluation of the methods, we found that the QE and BC sample mean estimators considerably outperformed the existing methods.

Although the BC sample standard deviation estimator performed best or comparably to the best performing method in the primary analyses of the simulation study, the sensitivity analyses and empirical evaluations did not clearly indicate a best performing approach for estimating the sample standard deviation. For all methods, the magnitude of the relative errors for estimating the sample standard deviation was typically higher than for estimating the sample mean.

In practice, the existing and proposed methods enable data analysts to incorporate studies that report medians in meta-analysis. Therefore, we compared the performance of the methods at the meta-analysis level using data from a real-life individual patient data meta-analysis. In this analysis, the methods that performed best for estimating the sample mean often resulted in the most accurate pooled mean estimates as well. As the QE and BC methods performed best for estimating the sample mean, these methods also performed best at the meta-analysis level.

In our empirical assessments, we assumed that all primary studies reported $S_1$, $S_2$, or $S_3$ summary data. Often in aggregate data meta-analyses, however, only a fraction of primary studies report $S_1$, $S_2$, or $S_3$ summary data and the other primary studies report sample means and standard deviations. Therefore, the results of our analyses at the meta-analysis level reflect the extremes in performance between the existing and proposed sample mean and standard deviation estimators. In practice, in meta-analyses where all or nearly all primary studies report medians, directly meta-analyzing medians may be better suited.[21, 33]

Notionally, the ABC and QE methods share numerous similarities and one may expect these methods to perform similarly to each other. In our analyses, three factors strongly differentiated the performance of these methods. First, the performance of ABC model selection was more highly variable and often favored the normal distribution (e.g., see simulation results for the Log-Normal(5, 0.25) distribution). Second, QE method gave more accurate estimates of the sample mean and standard deviation compared to the ABC method when data were not generated from one of the candidate parametric distributions. Finally, the ABC method was more sensitive to outliers. For example, the maximum values were highly variable when using the Log-Normal(5,1) distribution, and the method was highly biased in $S_1$ and $S_3$ even though the method correctly selected the log-normal distribution in nearly every repetition (e.g., see bottom row of Figure 1).

Our analyses focused on skewed data. As expected, when data were generated from a normal distribution, the Luo et al. sample mean estimators and the Wan et al. sample standard deviation estimators performed best (see Section 3 of Supplementary Material). However, most methods performed reasonably well in the normal case and the differences in performance amongst the methods were often inconsequential (e.g., AREs of magnitude less than 0.01 for the Luo et al., QE, and BC sample mean estimators in the Normal(5,1) case). When making the same assumption of normality when applying the QE or ABC methods (i.e., by only fitting the normal distribution), the performance of the methods improved but were still not superior to the Luo et al. and Wan et al. methods (data not shown).

This work has several limitations. Although the settings in our simulation study were based on those used in previous studies[13-17] to make a fair comparison between methods, these settings are not exhaustive and results may vary in other settings. Additionally, our simulation study focused solely on the performance of the methods for estimating the sample mean and standard deviation. In future work, we intend to conduct a simulation study investigating the performance of the methods at the meta-analysis level (e.g., for estimating the pooled effect measure and heterogeneity).

Strengths of this work include (*i*) comparing the recently developed Luo et al. method to the ABC method, (*ii*) including a greater number of outcome distributions compared to the simulation studies conducted by previous authors[13-15, 17], and (*iii*) empirically evaluating the accuracy of the methods using real-life data.

In summary, we recommend the QE and BC methods for estimating the sample mean and standard deviation when data are suspected to be non-normal, as they often outperformed the existing methods in the analyses presented herein. To make these methods widely accessible, we developed the R package 'estmeansd' (available on CRAN)[20] which implements these methods and launched a webpage (available at https://smcgrath.shinyapps.io/estmeansd/) that provides a graphical user interface for using these methods. We also encourage researchers performing meta-analysis to explore the sensitivity of their conclusions to the choice of method for estimating sample means and standard deviations.

**Declaration of conflicting interests**

The author(s) declared no potential conflicts of interest with respect to the research, authorship, and/or publication of this article.

**Table 1**: ARE of the methods when applied to estimate the sample means and standard deviations of the 58 primary studies. In each column, the ARE value closest to zero is in bold. The presented ARE values were rounded to two decimal places.

|         | ARE for $\bar{x}$ | | | ARE for $s_x$ | | |
|---------|-------|-------|-------|-------|-------|-------|
|         | $S_1$ | $S_2$ | $S_3$ | $S_1$ | $S_2$ | $S_3$ |
| Luo/Wan | -0.14 | -0.15 | -0.10 | -0.15 | **-0.01** | -0.08 |
| ABC     | -0.13 | 0.21  | -0.05 | -0.22 | 1.38  | -0.16 |
| QE      | **-0.05** | 0.06 | 0.00 | **-0.15** | 0.34 | **-0.08** |
| BC      | -0.08 | **0.00** | **0.00** | -0.25 | 0.06 | 0.11 |

**Table 2**: Estimates of the pooled mean PHQ-9 score and their 95% CIs when using the study-specific derived estimated sample means and standard deviations. For the pooled estimates under the "$S_1$", "$S_2$", and "$S_3$" columns, all methods were applied assuming $S_1$, $S_2$, and $S_3$ summary data, respectively, were extracted from all 58 primary studies, and the derived estimated study-specific sample means were meta-analyzed. When using the true study-specific sample means and standard deviations, the pooled estimate was 6.53 [95% CI: 5.97, 7.09]. In each column, the pooled estimate closest to the true value (i.e., 6.53) is in bold.

|         | $S_1$                  | $S_2$                  | $S_3$                  |
|---------|------------------------|------------------------|------------------------|
| Luo/Wan | 5.76 [5.15, 6.37]      | 5.68 [5.06, 6.29]      | 5.97 [5.36, 6.58]      |
| ABC     | 5.77 [5.13, 6.40]      | 7.12 [6.48, 7.77]      | 6.29 [5.69, 6.90]      |
| QE      | **6.26 [5.67, 6.85]**  | 6.88 [6.22, 7.53]      | **6.49 [5.92, 7.07]**  |
| BC      | 6.09 [5.48, 6.69]      | **6.59 [5.91, 7.28]**  | 6.58 [6.01, 7.14]      |

**Figure 1**: ARE of the Luo/Wan (red line, hollow circle), ABC (orange line, hollow triangle), QE (blue line, solid triangle), and BC (green line, solid circle) methods in scenario $S_1$. The panels in the left and right columns present the ARE of the sample mean estimators and sample standard deviation estimators, respectively.

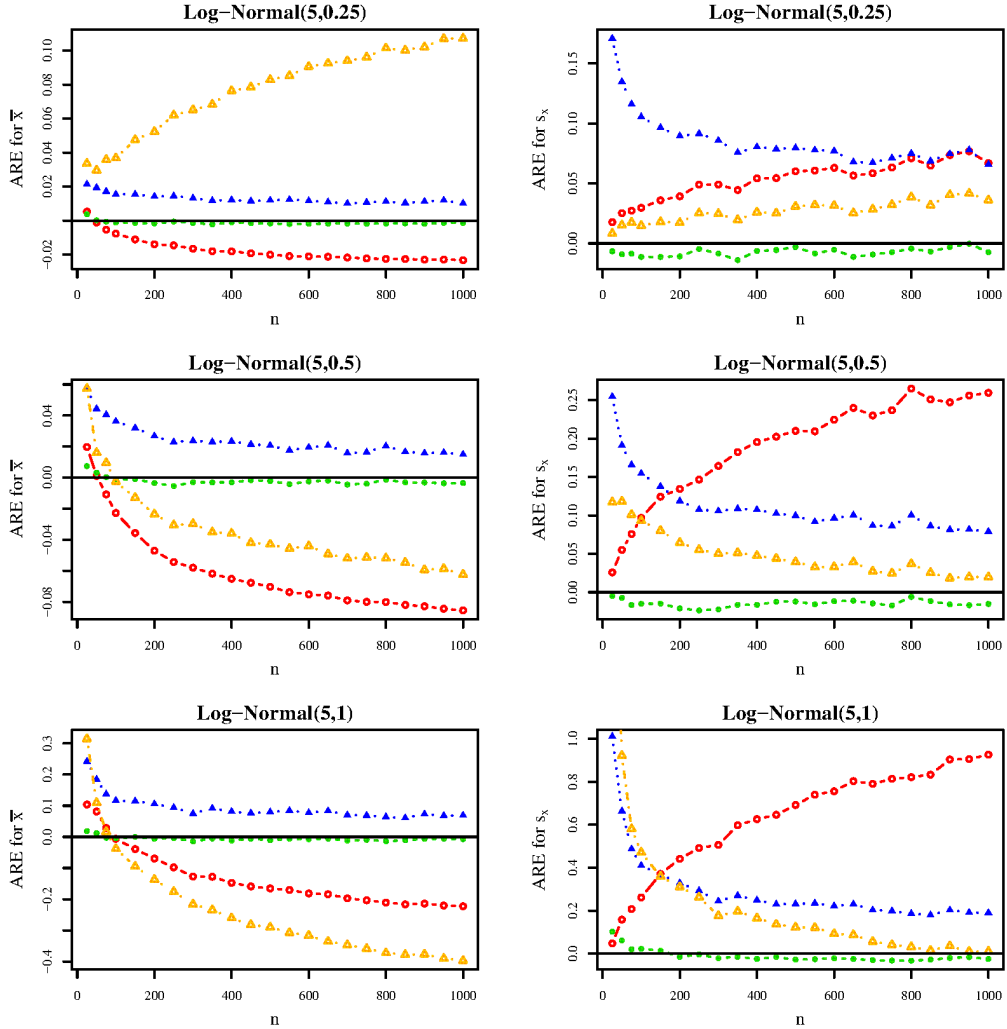

Note that for the Log-Normal(5,1) distribution, the ABC standard deviation estimator had ARE $= 2.05$ when $n = 25$.

**Figure 2**: ARE of the Luo/Wan (red line, hollow circle), ABC (orange line, hollow triangle), QE (blue line, solid triangle), and BC (green line, solid circle) methods in scenario $S_2$. The panels in the left and right columns present the ARE of the sample mean estimators and sample standard deviation estimators, respectively.

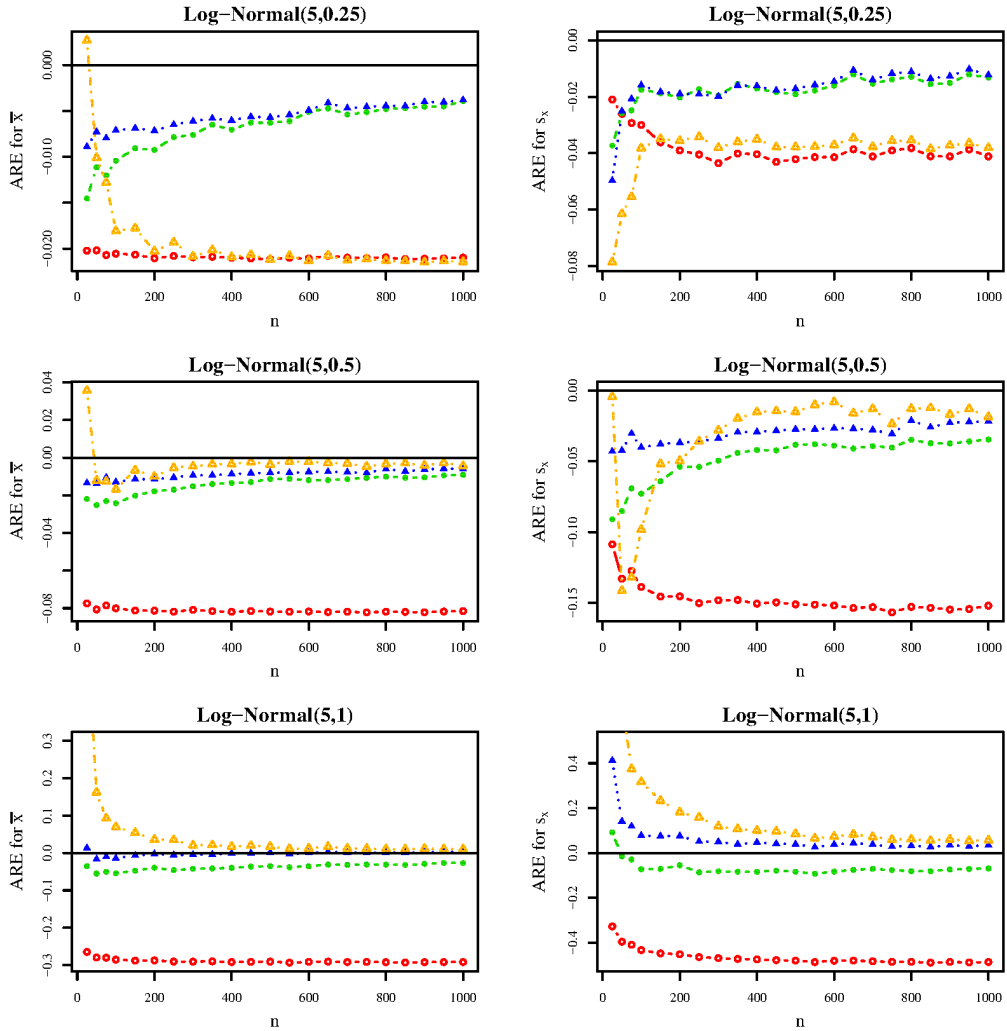

Note that for the Log-Normal(5,1) distribution, the ABC sample mean estimator had ARE = 0.59 when $n = 25$ and the ABC standard deviation estimator had ARE = 3.48 when $n = 25$ and ARE = 0.67 when $n = 50$.

**Figure 3**: Forest plot from the meta-analysis of mean PHQ-9 scores. The study-specific estimates represent the true sample means and their 95% CIs. The pooled estimate shown was obtained using the true-study-specific sample means and standard deviations. In the "Mean PHQ-9" column, the true study-specific sample means and their 95% CIs as well as the pooled mean and its 95% CI are given.

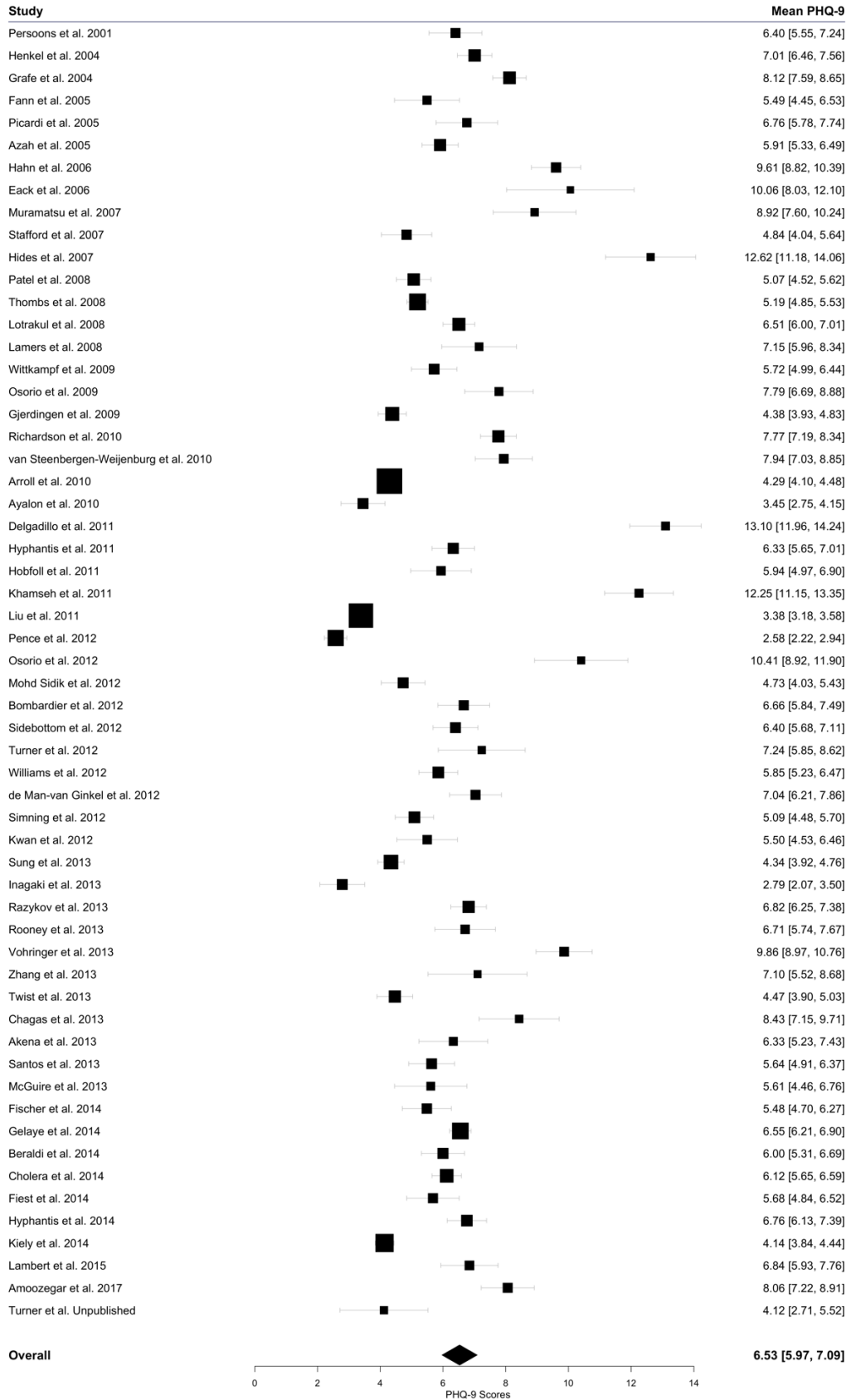

**Appendix A**

In the QE method, the parameters of a candidate distribution are estimated by minimizing the objective function, $S(\theta)$. This section describes the implementation of minimization algorithm.

We set the initial values for the parameters in the optimization algorithm as follows. First, we apply the methods of Luo et al.[17] and Wan et al.[15] to estimate the sample mean and standard deviation, respectively, from $S_1$, $S_2$, or $S_3$. Then, we apply the method of moments estimator of the candidate distribution using the estimated sample mean and standard deviation. The method of moments estimates of the parameters are used as the initial values of the parameters.

To minimize $S(\theta)$, we apply the limited-memory Broyden–Fletcher–Goldfarb–Shanno algorithm with box constraints (L-BFGS-B), which is implemented in the built-in 'optim' function in the statistical programming language R. Reasonable constraints for the parameters are imposed to improve the convergence of the algorithm (e.g., enforcing $\mu \in [Q_{\min}, Q_{\max}]$ for the Normal($\mu$, $\sigma^2$) distribution in $S_1$). The particular constraints are given in Table A1. These parameter constraints are based on the uniform prior bounds in the ABC method of Kwon and Reis[16]. In the simulation study, we found that the solution to the minimization problem was insensitive to perturbations of the parameter constraint values, provided the algorithm converged.

The algorithm is considered to converge when the objective function is reduced by a factor of less than $10^7$ of machine tolerance. In each application of the QE method in the simulation study, the algorithm converged for at least three distributions. If the algorithm failed to converge for a given candidate distribution, that candidate distribution was excluded from the model selection procedure.

**Table A1**: Parameter constraints for the L-BFGS-B algorithm.

| Scenario | Candidate Distribution | $\theta_1$ | $\theta_2$ |
| --- | --- | --- | --- |
| $S_1$ | Normal | $\mu \in (Q_{\min}, Q_{\max})$ | $\sigma \in (10^{-3}, 50)$ |
| | Log-Normal | $\mu \in (\log(Q_{\min}), \log(Q_{\max}))$ | $\sigma \in (10^{-3}, 50)$ |
| | Gamma | $\alpha \in (10^{-3}, 100)$ | $\beta \in (10^{-3}, 100)$ |
| | Beta | $\alpha \in (10^{-3}, 40)$ | $\beta \in (10^{-3}, 40)$ |
| | Weibull | $\lambda \in (10^{-3}, 100)$ | $k \in (10^{-3}, 100)$ |
| $S_2 \ \& \ S_3$ | Normal | $\mu \in (Q_1, Q_3)$ | $\sigma \in (10^{-3}, 50)$ |
| | Log-Normal | $\mu \in (\log(Q_1), \log(Q_3))$ | $\sigma \in (10^{-3}, 50)$ |
| | Gamma | $\alpha \in (10^{-3}, 100)$ | $\beta \in (10^{-3}, 100)$ |
| | Beta | $\alpha \in (10^{-3}, 40)$ | $\beta \in (10^{-3}, 40)$ |
| | Weibull | $\lambda \in (10^{-3}, 100)$ | $k \in (10^{-3}, 100)$ |

## Appendix B

To estimate sample mean and standard deviation using the BC method, the use of Box-Cox transformations requires the solutions to the following problems.

The first problem is defined as follows. In $S_1$, given $Q_{\min}$, $Q_2$, and $Q_{\max}$ such that $Q_{\min} < Q_2 < Q_{\max}$, find the finite power $\lambda$ of transformation such that

$$f_\lambda(Q_{\max}) - f_\lambda(Q_2) = f_\lambda(Q_2) - f_\lambda(Q_{\min})$$

Equivalently, this problem can be restated as finding $\lambda$ such that

$$\left(\frac{f_\lambda(Q_{\max}) - f_\lambda(Q_2)}{f_\lambda(Q_2) - f_\lambda(Q_{\min})} - 1\right)^2$$

is minimized to zero. Similarly, given $Q_1$, $Q_2$, and $Q_3$ such that $Q_1 < Q_2 < Q_3$, the corresponding minimization problem in $S_2$ is finding $\lambda$ such that

$$\left(\frac{f_\lambda(Q_3) - f_\lambda(Q_2)}{f_\lambda(Q_2) - f_\lambda(Q_1)} - 1\right)^2$$

is minimized to zero. Given $Q_{\min}$, $Q_1$, $Q_2$, $Q_3$, and $Q_{\max}$ such that $Q_{\min} < Q_2 < Q_{\max}$ and $Q_1 < Q_2 < Q_3$, the corresponding minimization problem in $S_3$ is finding $\lambda$ such that the following expression is minimized,

$$\left(\frac{f_\lambda(Q_3) - f_\lambda(Q_2)}{f_\lambda(Q_2) - f_\lambda(Q_1)} - 1\right)^2 + \left(\frac{f_\lambda(Q_{\max}) - f_\lambda(Q_2)}{f_\lambda(Q_2) - f_\lambda(Q_{\min})} - 1\right)^2.$$

To find $\lambda$, we use the built-in function 'optimize' in R. This function uses a combination of golden section search and successive parabolic interpolation for one-dimensional optimization.

The second problem arises when $\lambda < 0$ because in this case the mean and/or standard deviation are likely to be infinite. For example, $\lambda = -1$ results in a Cauchy distribution which has undefined mean and standard deviation. Therefore, we let $\lambda = 0$ in this case so that $\lambda$ is non-negative. By doing so, we implicitly assumed that the underlying distribution cannot be more heavy-tailed than a log-normal distribution. If this assumption does not hold, then estimating the mean and standard deviation of the underlying distribution may not be appropriate.

Supplementary Material for: Estimating the sample mean and standard deviation from commonly reported quantiles in meta-analysis

Sean McGrath, XiaoFei Zhao, Russell Steele, Brett D. Thombs, Andrea Benedetti and the DEPRESsion Screening Data (DEPRESSD) Collaboration

**Section 1**

In this section, we present the results of the sensitivity analyses of the simulation study for scenarios $S_1$ and $S_2$. Figures S1 and S2 give the $S_1$ and $S_2$ simulation results, respectively, for non-normal distributions.

**Figure S1**: ARE of the Luo/Wan (red line, hollow circle), ABC (orange line, hollow triangle), QE (blue line, solid triangle), and BC (green line, solid circle) methods in scenario $S_1$ in the sensitivity analyses. The panels in the left and right columns present the ARE of the sample mean estimators and sample standard deviation estimators, respectively.

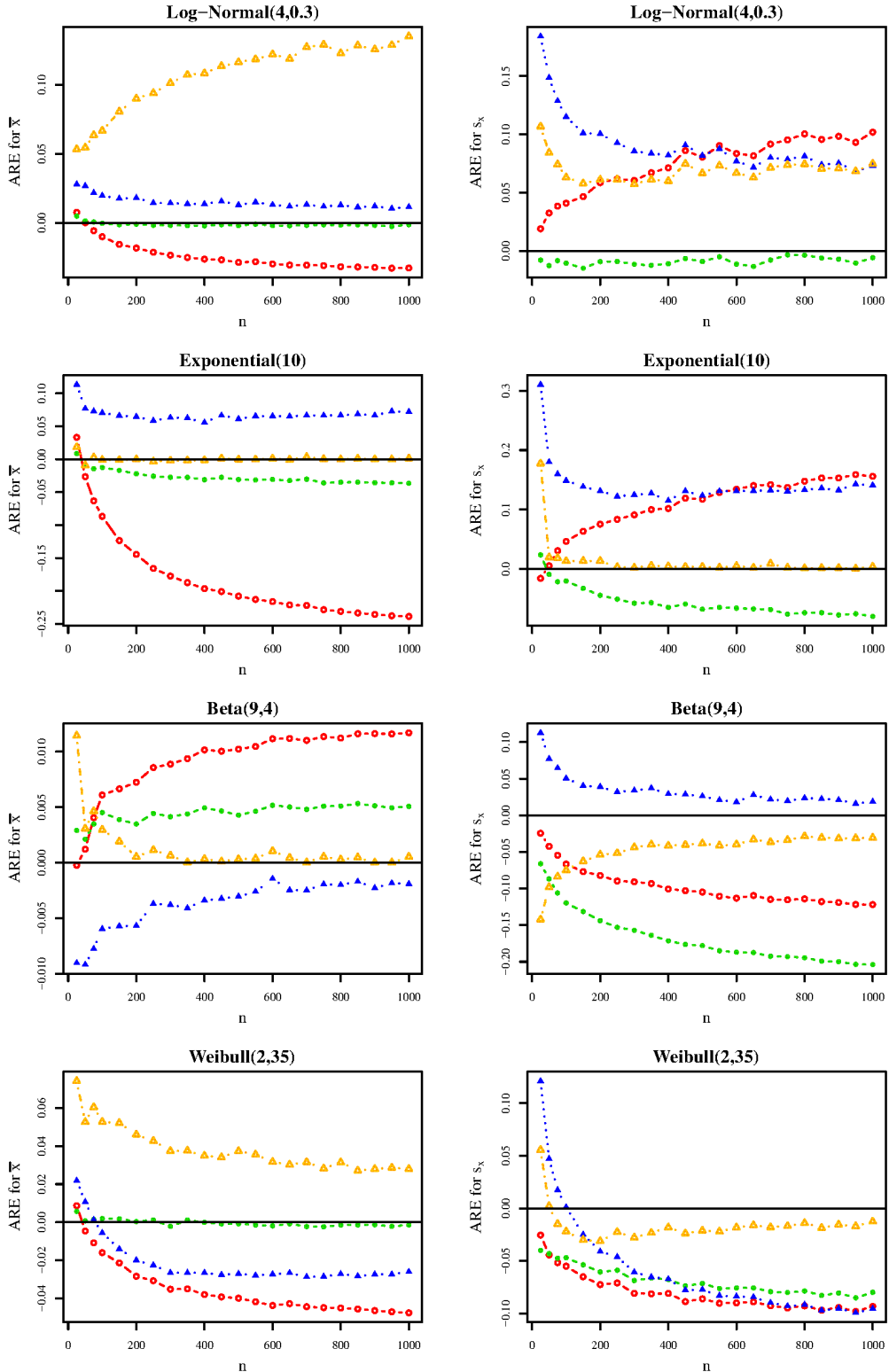

**Figure S2**: ARE of the Luo/Wan (red line, hollow circle), ABC (orange line, hollow triangle), QE (blue line, solid triangle), and BC (green line, solid circle) methods in scenario $S_2$ in the sensitivity analyses. The panels in the left and right columns present the ARE of the sample mean estimators and sample standard deviation estimators, respectively.

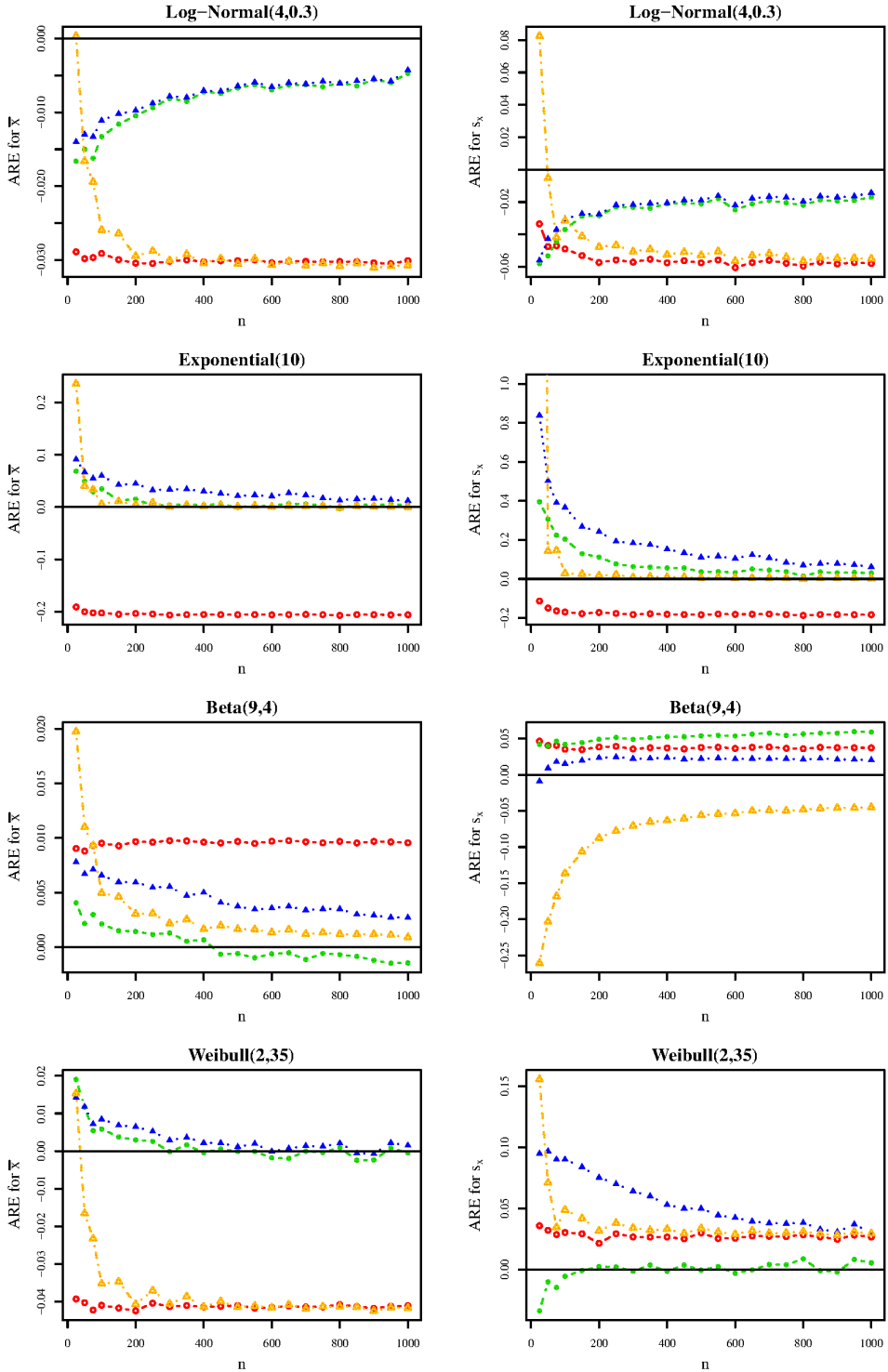

Note that for the Exponential(10) distribution, the ABC standard deviation estimator had ARE = 8.25 when $n = 25$.

**Section 2**

In this section, we present the $S_3$ simulation results. Figures S3 and S4 give the simulation results for the primary and sensitivity analyses, respectively.

**Figure S3**: ARE of the Luo/Wan (red line, hollow circle), ABC (orange line, hollow triangle), QE (blue line, solid triangle), and BC (green line, solid circle) methods in scenario $S_3$ in the primary analyses. The panels in the left and right columns present the ARE of the sample mean estimators and sample standard deviation estimators, respectively.

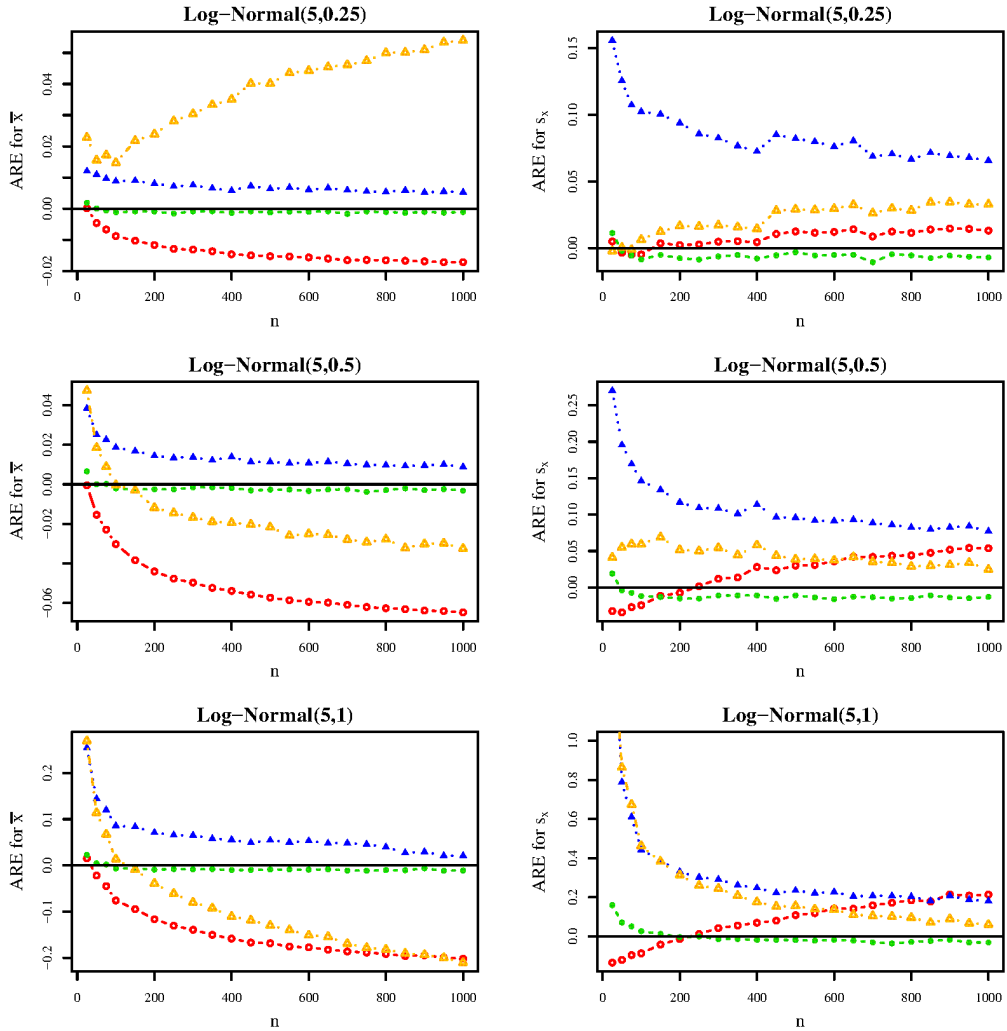

Note that for the Log-Normal(5,1) distribution, the QE and ABC standard deviation estimators had ARE = 1.70 and ARE = 1.57, respectively, when $n = 25$.

**Figure S4**: ARE of the Luo/Wan (red line, hollow circle), ABC (orange line, hollow triangle), QE (blue line, solid triangle), and BC (green line, solid circle) methods in scenario $S_3$ in the sensitivity analyses. The panels in the left and right columns present the ARE of the sample mean estimators and sample standard deviation estimators, respectively.

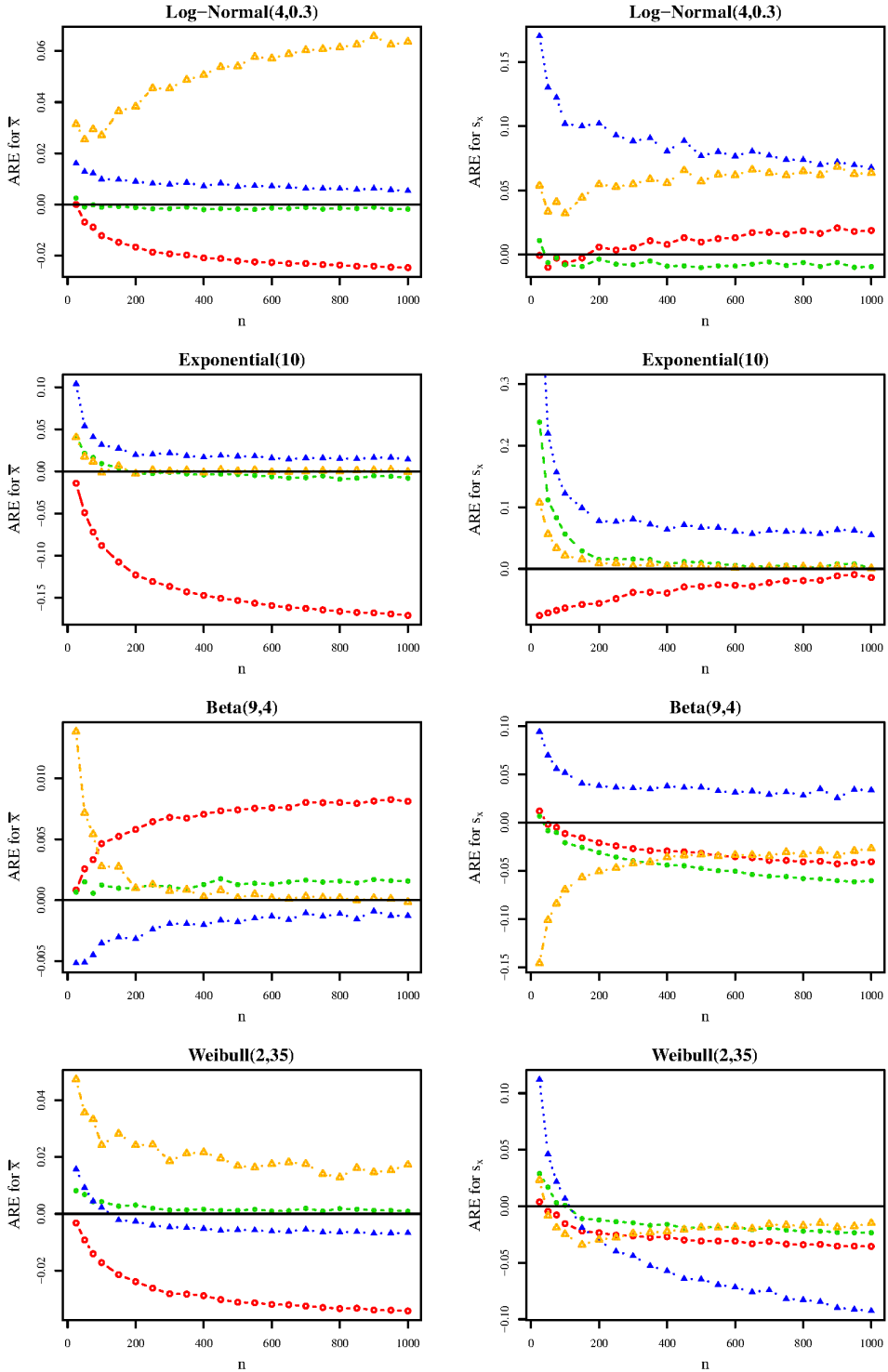

Note that for the Log-Normal(5,1) distribution, the QE standard deviation estimator had ARE = 0.51 when $n = 25$.

**Section 3**

In this section, we present the results of the simulation study when normal distributions were used to generate data. For these simulations, recall that the QE and ABC methods have candidate distributions including the normal distribution as well as several distributions with a strictly positive support. Therefore, a negative minimum value (in $S_1$ or $S_3$) or a negative first quartile value (in $S_2$) would bias QE and ABC model selection towards the normal distribution. Additionally, as described in the Example, the QE and ABC methods implicitly assume that the extracted summary data are strictly positive when fitting the log-normal distribution. Therefore, when applying all methods to data sampled from the normal distribution, if the extracted summary data included a negative value, the data were shifted so that the minimum value (in $S_1$ or $S_3$) or the first quartile value (in $S_2$) equaled 0.5. Let $c$ denote the value of such a shift. After estimating the sample mean, a value of $c$ was subtract from the sample mean.

Figures S5 and S6 give the simulation results for the primary and sensitivity analyses, respectively.

**Figure S5**: ARE of the Luo/Wan (red line, hollow circle), ABC (orange line, hollow triangle), QE (blue line, solid triangle), and BC (green line, solid circle) methods in scenario $S_1$ (top row), $S_2$ (middle row), and $S_3$ (bottom row) when applied to normally distributed data in the primary analyses. The panels in the left and right columns present the ARE of the sample mean estimators and sample standard deviation estimators, respectively.

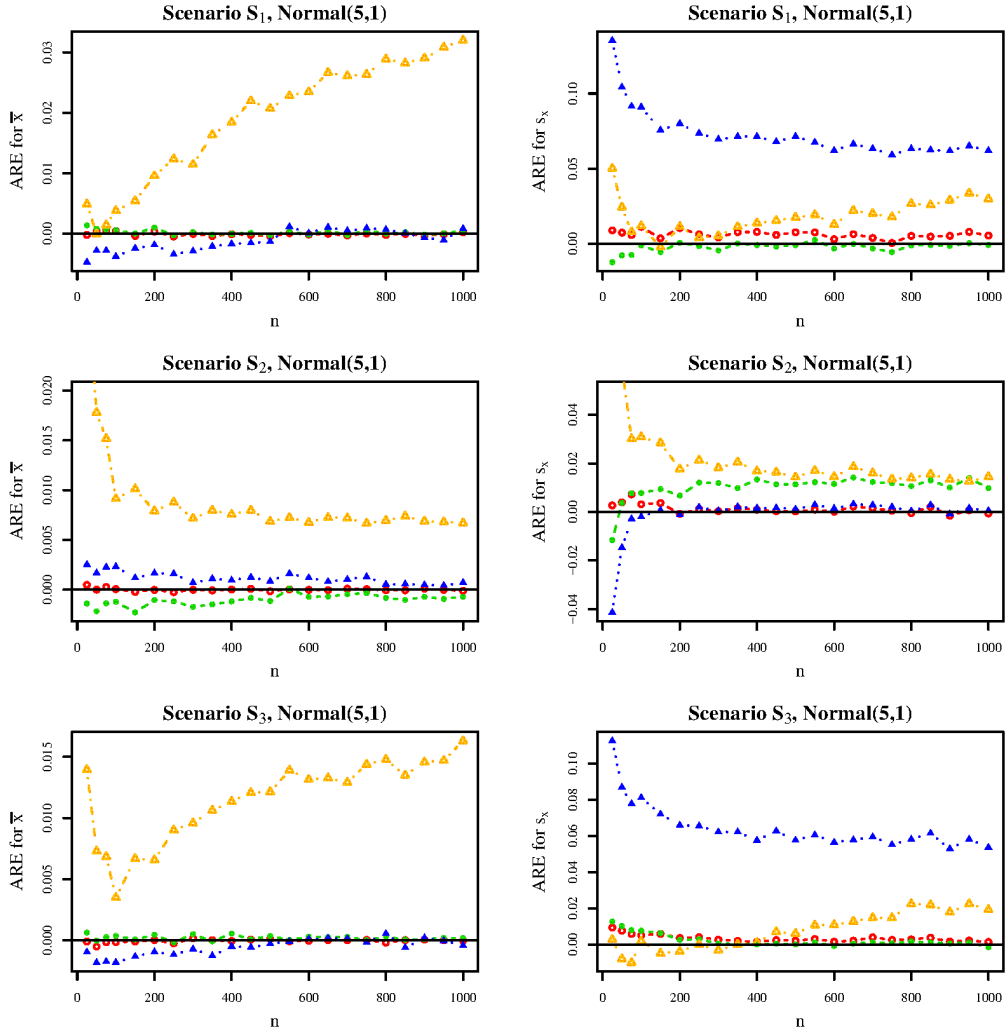

Note that in $S_2$, the ABC sample mean estimator had ARE = 0.03 when $n = 25$. Moreover, in $S_2$, the ABC standard deviation estimator had ARE = 0.18 when $n = 25$ and ARE = 0.06 when $n = 50$.

**Figure S6**: ARE of the Luo/Wan (red line, hollow circle), ABC (orange line, hollow triangle), QE (blue line, solid triangle), and BC (green line, solid circle) methods in scenario $S_1$ (top row), $S_2$ (middle row), and $S_3$ (bottom row) when applied to normally distributed data in the sensitivity analyses. The panels in the left and right columns present the ARE of the sample mean estimators and sample standard deviation estimators, respectively.

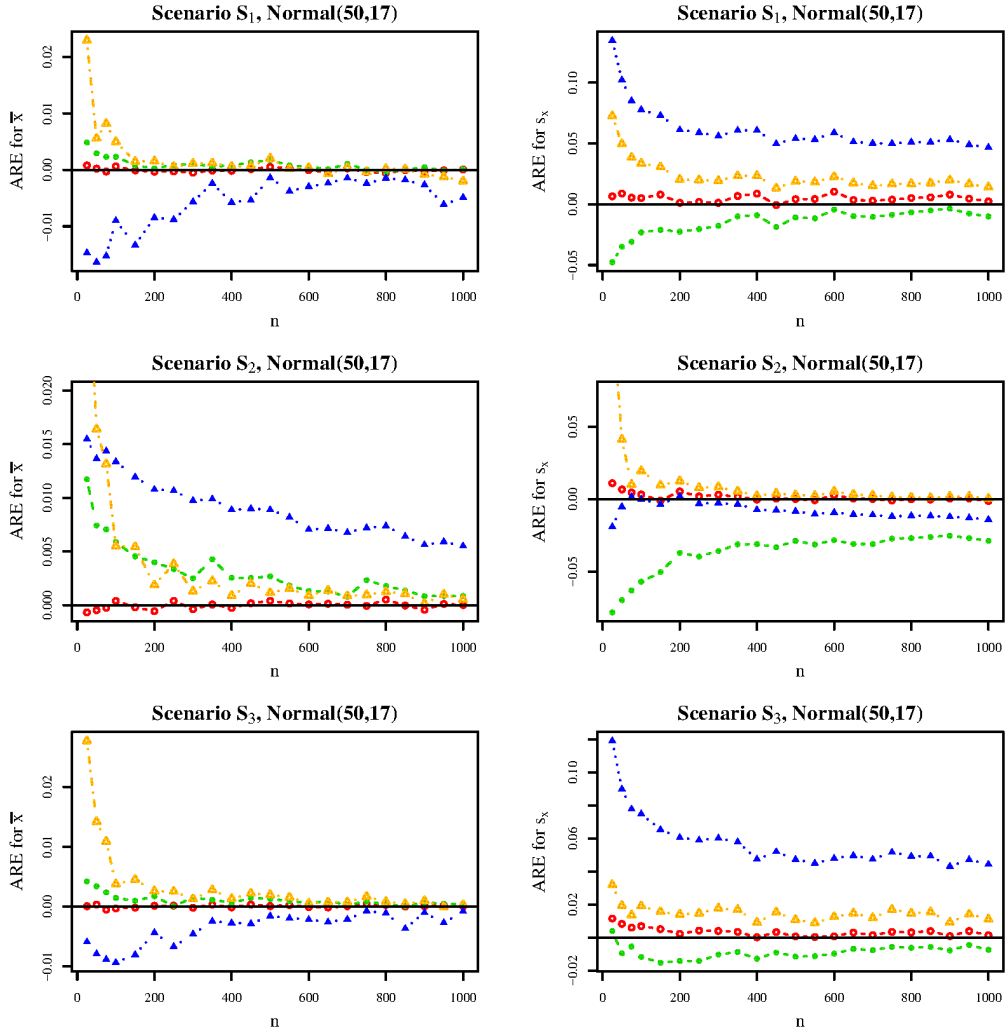

Note that in $S_2$, the ABC sample mean and standard deviation estimators had ARE = 0.03 and ARE = 0.13, respectively, when $n = 25$.

Section 4

**Table S1**: The sample minimum value ($Q_{min}$), first quartile ($Q_1$), median ($Q_2$), third quartile ($Q_3$), maximum value ($Q_{max}$), and sample size ($n$) of the 58 primary studies in the individual patient data meta-analysis of mean PHQ-9 scores.

| Study | $Q_{min}$ | $Q_1$ | $Q_2$ | $Q_3$ | $Q_{max}$ | $n$ |
|---|---|---|---|---|---|---|
| Persoons et al. 2001 | 0.00 | 2.00 | 5.00 | 9.00 | 27.00 | 173 |
| Henkel et al. 2004 | 0.00 | 3.00 | 5.00 | 10.00 | 25.00 | 430 |
| Grafe et al. 2004 | 0.00 | 3.00 | 7.00 | 12.00 | 27.00 | 494 |
| Fann et al. 2005 | 0.00 | 0.00 | 4.00 | 8.50 | 24.00 | 135 |
| Picardi et al. 2005 | 0.00 | 2.00 | 5.00 | 10.00 | 25.00 | 138 |
| Azah et al. 2005 | 0.00 | 3.00 | 5.00 | 8.00 | 21.00 | 180 |
| Hahn et al. 2006 | 0.00 | 5.50 | 9.00 | 14.00 | 26.00 | 211 |
| Eack et al. 2006 | 1.00 | 4.00 | 9.00 | 16.25 | 24.00 | 48 |
| Muramatsu et al. 2007 | 0.00 | 3.00 | 7.00 | 13.00 | 27.00 | 116 |
| Stafford et al. 2007 | 0.00 | 1.00 | 3.00 | 7.00 | 27.00 | 193 |
| Hides et al. 2007 | 0.00 | 6.00 | 13.00 | 18.50 | 27.00 | 103 |
| Patel et al. 2008 | 0.00 | 1.00 | 4.00 | 7.00 | 27.00 | 299 |
| Thombs et al. 2008 | 0.00 | 1.00 | 3.00 | 8.00 | 25.00 | 1006 |
| Lotrakul et al. 2008 | 0.00 | 3.00 | 6.00 | 9.00 | 24.00 | 278 |
| Lamers et al. 2008 | 0.00 | 3.00 | 5.00 | 12.00 | 27.00 | 104 |
| Wittkampf et al. 2009 | 0.00 | 1.00 | 4.00 | 9.00 | 27.00 | 260 |
| Osorio et al. 2009 | 0.00 | 1.00 | 5.00 | 14.00 | 24.00 | 177 |
| Gjerdingen et al. 2009 | 0.00 | 1.00 | 3.00 | 6.00 | 27.00 | 419 |
| Richardson et al. 2010 | 0.00 | 3.00 | 7.00 | 11.00 | 27.00 | 377 |
| van Steenbergen-Weijenburg et al. 2010 | 0.00 | 2.00 | 7.50 | 12.00 | 27.00 | 196 |
| Arroll et al. 2010 | 0.00 | 1.00 | 3.00 | 6.00 | 27.00 | 2528 |
| Ayalon et al. 2010 | 0.00 | 0.00 | 2.00 | 5.00 | 24.00 | 151 |
| Delgadillo et al. 2011 | 0.00 | 10.00 | 13.00 | 17.50 | 27.00 | 103 |
| Hyphantis et al. 2011 | 0.00 | 2.00 | 5.00 | 9.50 | 23.00 | 213 |
| Hobfoll et al. 2011 | 0.00 | 1.00 | 4.00 | 10.00 | 26.00 | 144 |
| Khamseh et al. 2011 | 0.00 | 6.00 | 11.00 | 19.00 | 27.00 | 184 |
| Liu et al. 2011 | 0.00 | 0.00 | 2.00 | 5.00 | 25.00 | 1532 |
| Pence et al. 2012 | 0.00 | 0.00 | 1.00 | 4.00 | 19.00 | 398 |
| Osorio et al. 2012 | 0.00 | 4.25 | 9.00 | 15.75 | 27.00 | 86 |
| Mohd Sidik et al. 2012 | 0.00 | 2.00 | 3.00 | 7.00 | 21.00 | 146 |
| Bombardier et al. 2012 | 0.00 | 2.00 | 5.00 | 10.00 | 27.00 | 160 |
| Sidebottom et al. 2012 | 0.00 | 2.00 | 5.00 | 9.00 | 26.00 | 246 |
| Turner et al. 2012 | 0.00 | 2.75 | 6.00 | 10.00 | 26.00 | 72 |
| Williams et al. 2012 | 0.00 | 2.00 | 5.00 | 8.00 | 21.00 | 235 |
| de Man-van Ginkel et al. 2012 | 0.00 | 3.00 | 6.00 | 10.00 | 23.00 | 164 |
| Simning et al. 2012 | 0.00 | 2.00 | 4.00 | 7.75 | 21.00 | 190 |
| Kwan et al. 2012 | 0.00 | 2.00 | 4.00 | 8.00 | 27.00 | 113 |
| Sung et al. 2013 | 0.00 | 1.00 | 3.00 | 6.00 | 27.00 | 399 |
| Inagaki et al. 2013 | 0.00 | 0.00 | 2.00 | 3.19 | 22.00 | 104 |

| Study | | | | | | |
|---|---|---|---|---|---|---|
| Razykov et al. 2013 | 0.00 | 3.00 | 6.00 | 10.00 | 26.00 | 345 |
| Rooney et al. 2013 | 0.00 | 3.00 | 5.00 | 9.00 | 25.00 | 126 |
| Vohringer et al. 2013 | 0.00 | 5.00 | 8.00 | 14.00 | 27.00 | 190 |
| Zhang et al. 2013 | 0.00 | 2.00 | 5.00 | 10.00 | 26.00 | 68 |
| Twist et al. 2013 | 0.00 | 0.00 | 2.00 | 7.00 | 27.00 | 360 |
| Chagas et al. 2013 | 0.00 | 4.00 | 7.50 | 12.00 | 23.00 | 84 |
| Akena et al. 2013 | 0.00 | 2.00 | 6.00 | 9.00 | 23.00 | 91 |
| Santos et al. 2013 | 0.00 | 1.00 | 4.00 | 8.00 | 21.00 | 196 |
| McGuire et al. 2013 | 0.00 | 1.00 | 4.00 | 8.50 | 23.00 | 100 |
| Fischer et al. 2014 | 0.00 | 1.00 | 4.00 | 8.00 | 27.00 | 194 |
| Gelaye et al. 2014 | 0.00 | 2.00 | 5.00 | 10.00 | 27.00 | 923 |
| Beraldi et al. 2014 | 0.00 | 3.00 | 6.00 | 8.00 | 16.00 | 116 |
| Cholera et al. 2014 | 0.00 | 2.00 | 5.00 | 9.00 | 22.00 | 397 |
| Fiest et al. 2014 | 0.00 | 1.00 | 4.00 | 9.00 | 26.00 | 169 |
| Hyphantis et al. 2014 | 0.00 | 2.00 | 5.00 | 10.00 | 27.00 | 349 |
| Kiely et al. 2014 | 0.00 | 1.00 | 3.00 | 6.00 | 27.00 | 822 |
| Lambert et al. 2015 | 0.00 | 2.00 | 6.00 | 10.00 | 24.00 | 147 |
| Amoozegar et al. 2017 | 0.00 | 3.00 | 7.00 | 12.00 | 27.00 | 203 |
| Turner et al. Unpublished | 0.00 | 0.50 | 3.00 | 5.00 | 24.00 | 51 |